\documentclass[aps,amssymb,amsmath,preprint, superscriptaddress, pra]{revtex4}
\usepackage{graphicx}
\usepackage{epsfig}
\usepackage{dcolumn}% Align table columns on decimal point
\usepackage{bm}% bold math
\usepackage{amsfonts}
\usepackage{color}
\usepackage{hyperref}
\usepackage{natbib}
\usepackage{epstopdf}
\usepackage[vcentermath]{youngtab}% for young tableaux
\usepackage{young}

\newcommand{\be}{\begin{equation}}
\newcommand{\ee}{\end{equation}}
\newcommand{\bc}{\begin{center}}
\newcommand{\ec}{\end{center}}
\newcommand{\bea}{\begin{eqnarray}}
\newcommand{\eea}{\end{eqnarray}}

\newcommand{\id}{1\hspace{-1.41mm}1}

\newcommand{\ket}[1]{|#1\rangle}

\definecolor{orange}{rgb}{1,0.5,0}
\begin{document}
\title{Qudit-Teleportation for photons with linear optics}

\author{Sandeep K. \surname{Goyal}}
%\email{goyal@ukzn.ac.za}
\affiliation{School of Chemistry and Physics, University of KwaZulu-Natal, Private Bag X54001, Durban 4000, South Africa}  
\affiliation{Optics and Quantum Information Group, The Institute of Mathematical Sciences, CIT campus, Chennai 600 113, India}

\author{Patricia E. \surname{Boukama-Dzoussi}}
%\email{boukfany@yahoo.com}
\affiliation{School of Chemistry and Physics, University of KwaZulu-Natal, Durban, South Africa}

\author{Sibasish \surname{Ghosh}}
%\email{sibasish@imsc.res.in}
\affiliation{Optics and Quantum Information Group, The Institute of
  Mathematical Sciences, CIT campus, Chennai 600 113, India} 
  
\author{Filippus S. \surname{Roux}}
%\email{fsroux@csir.co.za}
\affiliation{CSIR National Laser Centre, PO Box 395, Pretoria 0001, South Africa}

\author{Thomas \surname{Konrad}}
\email{konradt@ukzn.ac.za}
\affiliation{School of Chemistry and Physics, University of KwaZulu-Natal, Private Bag X54001, Durban 4000, South Africa}  
\affiliation{National Institute of Theoretical Physics (NITheP),  University of KwaZulu-Natal, Private Bag X54001, Durban 4000, South Africa}

%==================================================
\begin{abstract} 
Quantum Teleportation, the transfer of the state of one quantum system to another without direct interaction between both systems, is an important way to transmit information encoded in quantum states and to generate quantum correlations (entanglement) between remote quantum systems. So far, for photons, only superpositions of two distinguishable states (one ``qubit'') could be teleported.   Here we show how to  teleport a ``qudit'', i.e. a superposition of an arbitrary  number $d$ of distinguishable states  present in the orbital angular momentum of a single photon using  $d$ beam splitters and $d$ additional entangled photons. The same entanglement resource might also be employed to collectively teleport the state of $d/2$ photons at the cost of one additional entangled photon per qubit. This is superior to existing schemes for photonic qubits, which require an additional pair of entangled photons per qubit. 
\end{abstract}
%==================================================
\maketitle
%==================================================

In classical physics it is possible, in principle, to detect the state of a single system, for example the position and momentum of a point particle, transmit the information about that state to a remote location and then reconstruct it within a second system. This concept of ``classical teleportation'' underlies telecommunication techniques such as the transfer of documents via facsimile. Quantum physics, however, excludes the possibility to detect or duplicate the state of a single
microscopic system \cite{Wootters1982} and therefore  rules out all forms of classical teleportation with atoms, photons or other quantum
systems. It is thus surprising \cite{Werner1998}, that the state transfer   between
quantum systems can nevertheless be realized according to the rules of quantum
physics by means of ``quantum teleportation'' \cite{Bennett1993}. This procedure makes use of correlations between
quantum systems - entanglement - which cannot be described by
local-realistic theories \cite{Bell1964}, such as classical mechanics or
electrodynamics, nor any other theory within classical physics. 

Quantum teleportation lies at the core of quantum 
communication, which is the quantum analog of telecommunication, and can also be
employed to enhance the success probability in quantum computing with
photons \cite{Brassard1998,Gottesman1999,Knill2001,Duan2001}. Moreover, it is
one of the crucial ingredients \cite{Briegel1998, Lvovsky2009} for enabling long-distance quantum cryptography  - a technique to transmit information secured against eavesdropping.

The importance of quantum teleportation for quantum information processing and
communication can be seen from the long list of
experimental realizations of  teleportation of the state of a two-level
system corresponding to the smallest unit of quantum information - one quantum bit
(qubit) \cite{Bouwmeester1997,Boschi1998, Furusawa1998, 
  Nielsen1998, Pan2001, Jennewein2001, Kim2001, Lombardi2002, Marcikic2003,
  Bowen2003, Babichev2003, Ursin2004, Fattal2004}. In these realizations single qubits were encoded
in the polarization of photons or in the superposition of vacuum and
one photon states\cite{ Babichev2003}. Quantum teleportation with two-level atoms has been demonstrated
\cite{Riebe2004, Barrett2004, Olmschenk2009} and it has also been designed for three- and four-level  atomic systems \cite{Amri2010, Ritter2012}.

A way to identify quantum communication channels with high capacity (bandwidth), is to employ multi-level quantum systems that carry superpositions of an arbitrary number $d$ of basis states -- qudits --  instead of sending many single qubits. Due to their additional parameters, these systems might be optimized against the influence of external noise, such as decoherence of photonic states in turbulent atmosphere \cite{Hamadou.et.al13}. Moreover, it might save resources. For example, Quantum Key Distribution (QKD) schemes using photonic qudits can securely transmit more than one bit of information per photon \cite{Mafu.et.al13}. In order to achieve long-range QKD or quantum communication networks, so-called quantum repeaters  \cite{Briegel1998} that teleport qudits are obligatory. Instead of qudits one could teleport multiple qubits, but at the cost of additional transcription of qudits to qubits. While in theory the generalization of teleportation from qubits to qudits is known (cp.\ for example \cite{Werner01,Mor1999,Son2001,Gu2002,Kim2004}), the experimental realization of qudit teleportation is a difficult task and has not yet been achieved. 

At present, light is the only candidate for quantum
communication and quantum cryptography over large distances because of its small interaction with its environment as compared
to matter. %\footnote
The maximal distance for quantum communication achieved with
  photons so far is 144 km through the atmosphere \cite{Zeilinger07} limited mainly by
  absorption. Much further distances seem only possible using teleportation of
  entangled photons in conjunction with quantum repeaters.
%\footnote{The maximal distance for quantum communication achieved
%with photons so far was 144 km through the atmosphere limited mainly
%by absorption. Much further distances seem only possible
%using teleportation of entangled photons in conjunction with so-called
%quantum repeaters \cite{Briegel1998}}. 
At the same time, the small interaction of photons makes it difficult to manipulate the states of light in
order to achieve teleportation. There are two main challenges in realizing teleportation: (i)
photons sharing maximal quantum correlations 
(entanglement) have to be generated and distributed between the sender and the receiver of
quantum information and (ii) the input
photons and the photons of the sender have to be projected into a maximally entangled state by  a joint
measurement of both (a so-called Bell measurement) in order to transfer
the state of the input photons to the photons held by the receiver.   Both challenges can in principle be overcome using non-linear optical
media, which manipulate the light depending on its intensity. The corresponding processes, however, have a very
small efficiency on the {\em single photon level}. For example, in spontaneous parametric down-conversion, an incoming pump photon is converted into a pair of entangled photons [challenge (i)], with a probability of $10^{-6}$ per incoming photon \cite{Boyd}. The efficiency of a Bell measurement by means of  non-linear optics [challenge (ii)]  is even lower \cite{Kim2001}, at about  $10^{-10}$. Therefore it is desirable to design a more efficient solution to both challenges based on linear optics --- i.e. beam splitters, phase shifters, mirrors, etc. Such an efficient solution is presented in this article  for the teleportation of the quantum information carried by an arbitrary number $d$ of photonic levels. A different solution to challenge(ii) for qutrits ($d=3$) can  be found in \cite{Halevy2011}. An alternative teleportation scheme for general qudits based on quantum scissors together with a comparison to the present scheme is reported in \cite{Goyal2013}.

Although linear optics does not allow the realization of a complete Bell measurement \cite{Lutkenhaus1999}, a simple $50:50$ beam splitter can be used as a filter to project two incoming photons onto a  particular entangled state in a certain percentage of the cases. Two photons incident on the input ports of a beam splitter do not produce a coincidence count in two detectors in the output ports (Hong-Ou-Mandel effect \cite{Hong1987, Bose2002}) unless they possess an anti-symmetric component with respect to their internal degree of freedom, e.g.\ their polarization. A coincidence count thus effectively projects onto an antisymmetric state. For two polarized photons entering in different input ports of the beam splitter there is only one such state: 
\begin{align}
\ket{\psi}=& \frac{1}{\sqrt{2}}\left(\ket{H V}-\ket{V H}\right)\,,
\label{polanti}
\end{align} 
i.e.
elementary excitations of the first and second spatial mode (represented by the first and second slot
in the state symbol) carrying horizontal and  vertical  polarization, respectively, superposed with excitations of these modes with swapped polarizations.  This state is  antisymmetric, because it changes
sign under a permutation of the first and second mode
(slot), and  it is maximally entangled, a condition that allows the realization of the teleportation of a qubit \cite{Bouwmeester1997} encoded in the polarization of a single photon. Moreover, as we shall see in the following, this phenomenon can also be employed for the simultaneous teleportation of  multiple qubits encoded in the orbital angular momentum (OAM) of photons.

\begin{figure}
\includegraphics[width=0.3\textwidth]{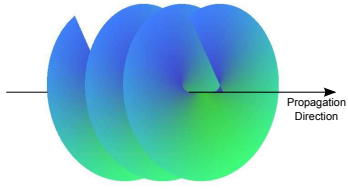}
\caption{Schematic diagram of the helical wavefront of a light beam.}\label{helical}
\end{figure}

It was noticed by Allen et al.\ \cite{Allen1992} in 1992 that light with a phase distribution
$\exp(i l \phi)$ depending on the azimuthal angle $\phi$ in the plane orthogonal
to its direction of propagation carries an orbital angular momentum of
an integer $l$ times Planck's constant $\hbar$ per photon. 
Such optical beams are characterized by helical (screw-like) wavefronts (see Fig.\ \ref{helical}) and can be produced with the aid of spatial light modulators (SLMs) --- thin liquid crystal displays. The SLM modulates the optical beam with a helical phase pattern.  
%Such light is characterized by helical (screw-like) wavefronts (cp. Fig.\ \ref{helical})  and can for example be produced by means of spatial light modulators - thin liquid crystal displays (LCDs)  which imprint the helical phase pattern or superpositions of such patterns.
The orbital angular momentum of a photon can thus be used  to carry information and represents  a quantum system with an unrestricted number of levels.

\begin{figure}
\includegraphics[width=8.5cm]{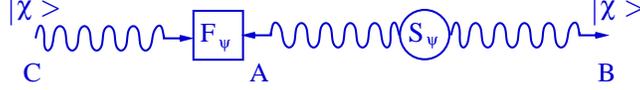}
\caption{Teleportation using a filter:  source $S_{\psi}$ produces a pair of systems $A$ and $B$ in state $\ket{\Psi}$. System $A$ and system $C$ are to be projected by a filter $F_{\psi}$ into state $\ket{\Psi}$. If successful, the filtering transfers the initial state of $C$ to system $B$.}\label{tele-schematic}
\end{figure}

Quantum teleportation using an incomplete Bell measurement ( a ``Bell filter'') can be applied to systems with an arbitrary number of levels, not only simple two-level systems (such as polarized photons).
Let us review how. Teleportation involves three parties Alice, Bob 
and Charlie, cp.\ Fig.~\ref{tele-schematic}. Alice and Bob are far apart and share a pair of systems in a 
maximally entangled state 
\begin{align}
|\Psi\rangle_{AB}
= &\frac{1}{\sqrt{D}}\sum_{i=0}^{D-1}|
A_i\rangle_A\otimes|B_i\rangle_B\,,
\label{Psi}
\end{align}
 which is a superposition of products of orthogonal
basis states of their $D$-level systems $A$ and $B$. Charlie
provides Alice with  an unknown state $|\chi\rangle$, which has to be
transferred from Charlie's system $C$
to Bob's system $B$. 
For this purpose systems $C$ and $B$  must be
similar - $B$ has to support the same states as $C$. The state to be
teleported can thus be expressed as a superposition of Bob's basis states:
$|\chi\rangle=\sum_{k=0}^{D-1}\alpha_k|B_k\rangle$. 
Alice successfully teleports   
the state
$|\chi\rangle$ if she carries out a measurement that acts like a filter and projects systems
$C$ and $A$ onto the entangled state $|\Psi\rangle$: 

\begin{align}
|\chi\rangle_C\otimes|\Psi\rangle_{AB} \rightarrow &
\left(\left[|\Psi\rangle\langle\Psi|\right]_{CA}  \otimes \id_B
 \right)
|\chi\rangle_C\otimes|\Psi\rangle_{AB} \nonumber \\
& = \frac{1}{D^{3/2}}\sum_{k,l= 0}^{D-1} \alpha_k
|B_l\rangle_C\otimes|
A_l\rangle_A \otimes |B_k\rangle_B  \nonumber \\ 
%\langle B_m|B_i\rangle_C\langle A_m|A_k\rangle_A 
& = \frac{1}{D}|\Psi\rangle_{CA}\otimes |\chi\rangle_B\,.
\label{teleport1}
\end{align}
According to the rules of quantum mechanics the likelihood for such a
projection to occur is given by the  square of the
length of the resulting state vector, $p=1/D^2$. Thus the success probability 
of this teleportation scheme, which uses only one of the outcomes of
a Bell measurement, decreases with the number $D$ of basis
states in which quantum information is encoded. The advantage lies in the fact that this concept of teleportation, which is used for photonic qubits ($D=2)$  can be generalised to teleport states with arbitrary $D$ using linear
optics. %\footnote
({For an alternative approach based on quantum scissors \cite{Pegg1998}  that uses  entanglement contained in single-photon states see \cite{Goyal2013}}). 

\section*{Results}
To generalize the qubit scheme to a method that is able to teleport superpositions of multiple basis states, we have to
identify a photonic system with a unique antisymmetric state and a linear
optical device that plays the role of the beam splitter in the qubit
case, i.e. a filter for antisymmetric states.   The uniqueness is required to guarantee that the filter yields the same state $\ket{\Psi}_{CA}$ that is initially shared by Alice and Bob, i.e. the state $\ket{\Psi}_{AB}$.     The dimension of the space spanned by the antisymmetric states of composite systems (only if the dimension equals $1$ do we have a unique antisymmetric state!) can be easily determined by means of Young tableaux (see Methods).  It turns out that only $d$ systems each with $d$ levels posses a unique antisymmetric state. As a consequence, a generalization of the teleportation scheme for photonic qubits by means of a Bell filter for antisymmetric states requires $d$ photons propagating on different paths each with a quantized degree of freedom, such as OAM with $d$-levels. In other words, the generalization requires $d$ qudits.

Photonic states can be conveniently expressed by means of creation
operators $a^\dagger$ acting on the vacuum state $\ket{0}$. In our case these operators carry two indices ---
the first one, $j$, specifies one of $d$ possible propagation paths whereas the
second one, $l=1\ldots d$, denotes the OAM value
$l\hbar$ of the photon. For example, the state $\ket{\psi}= (\ket{12}- \ket{21})/\sqrt{2}$ of two
photons propagating
on different paths with two OAM values $l=1,2$ --  which is the OAM analog of the antisymmetric polarization
state \eqref{polanti} used in qubit teleportation  \cite{Bouwmeester1997} -- can be written by means of a determinant of
creation operators:
\begin{align}
\ket{\psi}= &\frac{1}{\sqrt{2}}\left( a_{11}^\dagger a_{22}^\dagger - a_{12}^\dagger
a_{21}^\dagger \right) \ket{0}= \frac{1}{\sqrt{2}}\det \left(\begin{array}{cc}
a_{11}^{\dagger}& a_{12}^{\dagger}\\
a_{21}^{\dagger}& a_{22}^{\dagger}\\
\end{array}\right)\ket{0}.
\end{align}
It is obvious that the state $\ket{\psi}$ is antisymmetric, since under
permutation of the propagation paths it is transferred to
$-\ket{\psi}$. The antisymmetry is represented by the determinant: a
swap of rows corresponding to the permutation results in a minus sign
of the determinant. Using the same logic, one can show that 
the antisymmetric state of $d$ photons with $d$ OAM values can be
expressed by the determinant of a $d\times d$ matrix $\Lambda$ 
\begin{align}
|\Psi\rangle &= \frac{1}{\sqrt{d!}}\det(\Lambda)|0\rangle\,
\label{antisym}
\end{align}
where
\begin{align}
\Lambda &= \left(\begin{array}{cccc}
a_{11}^{\dagger}& a_{12}^{\dagger}& \cdots & a_{1d}^{\dagger}\\
a_{21}^{\dagger}& a_{22}^{\dagger}& \cdots & a_{2d}^{\dagger}\\
\vdots & \vdots & \ddots & \vdots \\
a_{d1}^{\dagger}& a_{d2}^{\dagger}& \cdots & a_{dd}^{\dagger}
\end{array}\right).\label{lambda}
\end{align}
Here $\ket{\Psi}$ is the required antisymmetric state. The antisymmetry  follows from the fact that permutation of any two propagation directions corresponding to a
swap of two rows of the determinant in Eq.\ (\ref{antisym}) leads to a change of sign of the state $\ket{\Psi}$. 
\begin{figure}
\includegraphics[height=0.15\textwidth]{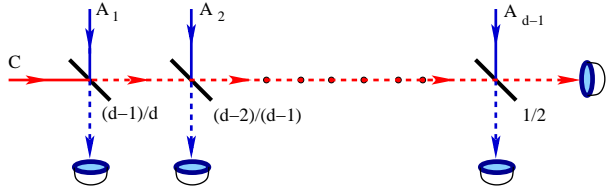}
\caption{Bell filter: an array of $d$ beam splitters (with indicated transmitivities) projects Charlie's incoming photon and Alice's $d-1$  photons into the antisymmetric state $\ket{\Psi}$ in case of a coincidence count in all detectors. For this purpose Alice's photons must be already in an antisymmetric state before the projection. To take care of the change of sign of OAM values upon reflection at the beam splitters, the OAM values carried by the light in the horizontal path must have opposite sign to those in the vertical paths.}\label{multi-bs}
\end{figure}  

For teleportation the $d$ photons in the antisymmetric state given in Eq.\ (\ref{antisym}) must be divided between Alice ($n$ photons) and Bob ($d-n$
photons) such that both share a bipartite maximally entangled state. 
To check for maximal entanglement we 
take advantage of the representation of $\ket{\Psi}$ in terms of a
determinant. Expanding the determinant with respect to the first row,
we obtain the state: 
\begin{align}
\ket{\Psi}=&\frac{1}{\sqrt{d!}}\det(\Lambda)|0\rangle
   =\frac{1}{\sqrt{d!}}\sum_{i=1}^d (-1)^{i+1} a^\dagger_{1i} \det
   (\Lambda_{1i})|0\rangle \nonumber \\
   = &\frac{1}{\sqrt{d}}\sum_{i=1}^d \ket{A_i}\ket{i}\,,
\end{align}
where $\ket{A_i}=1/\sqrt{(d-1)!}\det
   (\Lambda_{1i})|0\rangle$ and $\Lambda_{1i}$ is the $(d-1)\times(d-1)$ submatrix obtained by
   omitting the $i$-th column and the first row (a so-called minor of $\Lambda$) and $(-1)^{i+1} a_{1i}^\dagger|0\rangle = |i\rangle$. It is remarkable that the expansion of the determinant results in a
   maximally entangled bipartite state of the form (\ref{Psi}) with
   $D=d$ and $\ket{B_i}=\ket{i}$, implying that
   Alice and Bob obtain $d-1$ and one photon, respectively. 
This partitioning of photons allows Alice to teleport any state of a single $d$-level photon from Charlie by sending it along with her $d-1$ photons into the beam splitter array depicted in Fig.\ \ref{multi-bs} and subsequently obtaining a coincidence count in all its output ports. The coincidence count results effectively in a projection into the antisymmetric state $\ket{\Psi}$ as can be checked for any finite dimension $d$ by direct calculation. 
On the other hand, the antisymmetric state entering the beam splitter array  yields, with certainty, a coincidence count (cp. Methods).

\begin{figure}
\includegraphics[width=7.5cm]{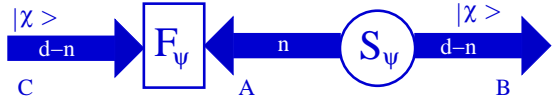}
\caption{Teleportation of the collective antisymmetric state  $\ket{\chi}$ of $d-n$ photons (number of photons indicated).}\label{fig-03}
\end{figure}

\begin{figure}
\includegraphics[width=7.5cm]{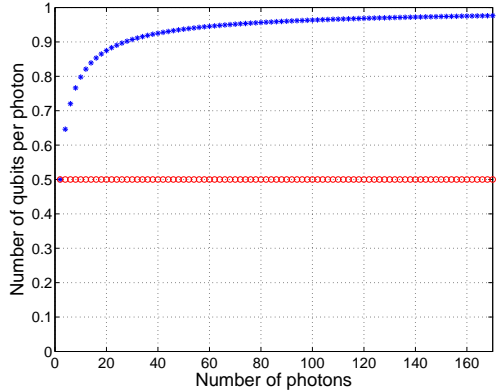}
\caption{The graphs show the number of quantum information units (qubits) teleported per additional photon versus the number of additional photons required  for  individual-qubit teleportation (red) and the optimal new  technique (blue). }\label{n-vs-qubits}
\end{figure}

But this is not the only possible partitioning of photons that leads to a
maximally entangled state between Alice and Bob. Strikingly, any partition
$(n, d-n)$ with $0<n< d$ of an antisymmetric state of $d$ particles possesses this property, and this can be easily understood by virtue
of the rules to calculate determinants (see Methods). For the partition 
$(n, d-n)$ one obtains a bipartite state as given in Eq.\ (\ref{Psi}) with $D= \binom{d}{n}$ and $\ket{A_i}$ as well as
$\ket{B_i}$ given in terms of minors of $\Lambda$. 
The maximum amount of quantum information is teleported with a $(d/2, d/2)$ partition of an even number $d$ of photons prepared in state $\ket{\Psi}$. In this case Charlie has to provide $d/2$ photons, cp.\ Fig.~\ref{fig-03}, and can send for large $d$ approximately $d$ qubits, simultaneously (cp.\ Fig.~\ref{n-vs-qubits}). This information  is encoded in the $D= \binom{d}{d/2}= 2^{d} + O(d)$ dimensional, anti-symmetric subspace  spanned by the $\ket{B_i}$.  This is an exponential gain compared to the teleportation of a single qudit (corresponding to $\log_2(d)$ qubits)  carried by one photon, which requires the same entanglement resource -- an antisymmetric state of $d$ photons $\ket{\Psi}$. %\footnote
({Note that Charlie's photons must be in an antisymmetric state to guarantee that the beam splitter array acts as a Bell filter in case of a coincidence count in its output ports. A general Bell filter for $n$ photons from Alice and $d-n$ photons from Charlie in antisymmetric states can be constructed by the recursive use of the Bell filter depicted in  Fig.~\ref{fig-03} combined with heralding techniques for single photons, cp.\ Fig.~\ref{general}. An example of a Bell-filter without heralding techniques for qutrits is discussed together with the preparation of a totally antisymmetric state in Methods.}).  Let us emphasize that this method thus constitutes a simultaneous teleportation of the collective state of several photons.     

\section*{Discussion}
In comparison, teleporting $d$ qubits individually, one requires the same number of maximally entangled photon pairs, i.e. a total of $2 d$ additional photons, resulting in an efficiency of sending half a qubit per additional photon or $1/3$ of a qubit per photon generated. Our scheme yields double these rates and thus requires half the number of photons (cp Fig.~\eqref{n-vs-qubits}) to teleport the same  amount of quantum information. This is an important improvement since the number of photons, that can be generated per time-unit is the limiting factor for the bandwidth of photonic quantum communication. However, it holds only for the selective regime conditioned on successful teleportation events. On average the number of photons increases by a factor $D^2$ with the number  $D=d^d$ of  outcomes of a Bell measurement of $d$ qudit photons. Since it might be possible to create Bell filters by means of linear optics for half of the Bell states (cp.\ \cite{Lutkenhaus1999}) instead of a single filter as in our case, this would mean an increase of only a factor $2$. This is a promising subject for future research since for individual qubit teleportation with linear optics the success probability is limited by $1/2^d$, resulting in a relative exponential overhead.    

In the foregoing presentation of results we assumed the existence of a source $S_\Psi$ that produces the state $\ket{\Psi}$ initially shared by Alice and Bob (cp.\ challenge (i)) which is treated in the Methods. Indeed for qutrits, such a source can be realized by an additional beam splitter array with two beam splitters (cp.\ Fig.\ \ref{multi-bs} for $d=3$),  that acts as a Bell-filter for the antisymmetric state $\ket{\Psi}$. For this purpose a single photon together with an antisymmetric state of two photons must enter the array and one photon must leave each of its output ports.
%This method together with one to produce the source for photonic quadit teleportation is discussed in the Methods. 
In the general case nondestructive heralding techniques based on non-linear optical effects \cite{Konrad2006} might be employed in conjunction with photonic multiports, which however reduce the efficiency of the teleportation scheme. These methods 
%with a scheme based on quantum scissors \cite{Pegg1998} 
will be discussed elsewhere.   

Apart from an additional overhead of photonic resources needed if filtering techniques are used to prepare the antisymmetric state (cp.\ first paragraph in this section) there is the need to produce simultaneously the number of photons required for the entangled state and the state that  carries the quantum information to be teleported. If the photons are generated by single photon sources, the success probabilities for the single photons will multiply and lead to an exponential decrease of the total success probability. This is a serious problem, however, it is also encountered when qubits in an unknown state that cannot be resent are to be teleported individually. To solve this problem one requires techniques investigated for long distance quantum communication and require to store the required number of photons in quantum memories and release them collectively to produce the entangled states needed for the teleportation of complex states.

 \section*{Methods}

 \noindent {\bf Dimension of the antisymmetric subspace}\\
The dimension of an antisymmetric subspace can be calculated using combinatorial objects called Young tableaux, which provide a technique of keeping track of the constraints imposed by the permutation symmetry of the system. Here we represent a basis state of a system by a box, $\young(a)$, where $a$ numbers the basis state. A basis of the symmetric combinations of two systems can be depicted by a row of two boxes,  {\small$ \yng(2)$}. Similarly, a basis of anti-symmetric states is represented by a column of boxes, {\tiny$\yng(1,1)$}. Since we are interested only in the antisymmetric part, we focus on  columns only. The dimension of the corresponding subspace, i.e. the number of basis states,  is obtained by counting the different possible ways to  fill the boxes with numbers according to certain rules. For the antisymmetric subspace we start filling the numbers in descending order, from top to bottom. For a system that consists of two subsystems, a Young tableaux reads:
\begin{align}
\young(a,b) \equiv |a\,b\rangle - |b\,a\rangle,
\end{align}
where $a$ is always greater than $b$. Therefore, if the total number of basis states available for each subsystem is two, i.e. we are dealing with two qubits, there is only one possibility, namely $a=2$ and  $b=1$. As a result, we obtain an antisymmetric subspace of dimension one. If the available states are more than two, say $d$, then we have $d-1$ options for $a$ and given $a$,  $a-1$ options for $b$. As a result the total number of combinations is given by $1 + 2 + \ldots + d-1=d(d-1)/2$, which is the dimension of the antisymmetric subspace for a pair of $d$-level systems, each of which can carry one qudit of quantum information. 

This can be generalized for systems with $n$ subsystems. Now we have $n$ numbers $\{a_1>a_2>\cdots >a_n\}$ in a column of boxes. The dimension of this antisymmetric subspace is given by the binomial coefficient ${\small \left(\begin{array}{c}d\\n\end{array}\right)}$, i.e. $d$ choose $n$, where $n$ is the number of subsystems. This equals $1$ only when $n=d$, giving us a unique antisymmetric state for $d$ qudits.\\

\noindent {\bf Laplace expansion for the determinant of a matrix}\\
The determinant of a  $n\times n$ matrix $A$ with elements $a_{ij}$ can be calculated by an expansion with respect to the first row of $A$ as follows:
\begin{align}
\det(A) &= \sum_{i=1}^n (-1)^{1+i} a_{1\, i}\det(A_{1\, i}).
\end{align}
Here $\det(A_{1\, i})$ is the determinant of the $(n-1)\times (n-1)$ submatrix of $A$ obtained from eliminating the first row and $i$-th column. In fact this is just a special case of a simultaneous expansion of the determinant with respect to several rows. For example, expanding with respect to the first two rows of a $4\times 4$ matrix $A$, we obtain:
%\begin{center}
\begin{eqnarray}
\includegraphics[width=8.5cm]{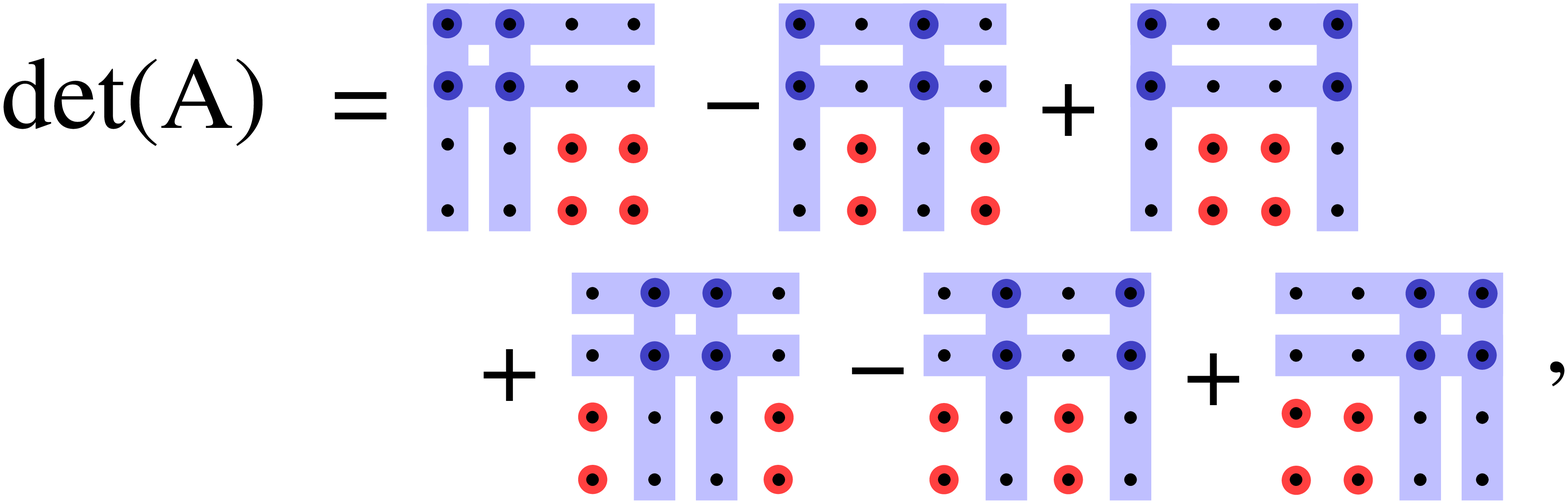}
\end{eqnarray}
%\end{center}
where each  block on the right-hand side represents the product of the determinant of the submatrix (minor) with blue dot and the minor with the red dot. In general, any such Laplace expansion assumes the form $\det(A)=\sum_i c_i \det(A_i)\det(B_i)$, where $c_i=\pm1$ and the $A_i$ ($B_i$) are minors of $A$  which differ at least in one column \cite{Lancaster}. The possible Laplace expansions of $\det{\Lambda}$ in Eq.~(4) %\ref{antisym}) 
correspond to the different distributions of the $d$ photons between Alice and Bob.  Each distribution leads to orthogonal states $\ket{A_i}\propto \det(A_i)\ket{0}$ on Alice's side, and on Bob's  side accordingly, and therefore to a maximally entangled state shared between both parties.
\\
%\noindent {\bf Antisymmetric state induces coincidence count in multiport}\\
%An antisymmetric state of $d$ qudits entering a beam splitter arrangement with $d$ input and $d$ output ports (multiport) results in a coincidence count in detectors in the output ports.  Since the beam splitters only superimpose the incoming light fields (first index of the
%creation operators)  independent of their orbital angular
%momenta (second index), the resulting linear transformation can be expressed
%by a rotation matrix $U$ acting on the creation operators    
%\begin{align}
%\tilde{a}_{kj}^{\dagger} &= \sum U_{ki}a_{ij}^{\dagger}\,,
%\end{align}
%which leads to a linear transformation of the matrix $\Lambda$
%\begin{align}
%\tilde{\Lambda} &= U\Lambda
%\end{align}
%and the invariance of the antisymmetric state: $|\Psi\rangle \to
%\det(\tilde{\Lambda})|0\rangle =
%\det(U)\det(\Lambda)|0\rangle = \det(\Lambda)|0\rangle=|\Psi\rangle $. 
%Hence one photon will leave from  each of the $d$ output ports of the beam splitter array
%thus leading to a possible coincidence count.\\
%\\
\noindent {\bf Proof of Bell Filter projection onto the antisymmetric state}\\ 
The goal of this section is to prove that the Bell filter depicted in Fig.~\eqref{multi-bs} projects onto the totally antisymmetric state of $d$ photons upon simultaneous detection in all $d$ detectors of the filter. We assume that the input is given by the combined state of Alice's  $d-1$ photons, which are in an antisymmetric state,  and  Charlie's single photon. We will consider all the possible states obeying this assumption and show that only the totally antisymmetric state results in a coincidence count.  We assume in the following that Charlie's single photon is prepared as a superposition of basis states with the same modulus of OAM values as used to span the Hilbert space for Alice's photons,  but with different signs: $-l_{i_C}=l_{i_B}$ for $i=1\dots d$. With this sign convention we can ignore the effect of sign change of the OAM modes ($l\rightarrow -l$) upon reflection on a beam splitter or mirror.

Note that $d-1$ photons each with $d$ levels form a  $d$-dimensional antisymmetric subspace of the $d^{d-1}$-dimensional Hilbert space. Therefore, the effective dimension of the allowed state space for Alice's $d-1$ photons is the same as the dimension of Charlie's Hilbert space which corresponds to a  single photon with $d$ levels.

The general quantum teleportation protocol for $d$-level systems rests on the complete projective measurement in the maximally entangled generalized Bell basis $\{\ket{\Psi}_{m_1m_2}\}$, which can be defined as:
\begin{align}
|\Psi\rangle_{m_1m_2} &= \sum_{n=0}^{d-1} \omega^{nm_1} \ket{A_n} \otimes  \ket{B_{n+m_2}}
\end{align}
where $m_1,m_2 = 0,1,\cdots,d-1$ and $\omega = \exp(2\pi inm_1/d)$. For $m_1 = m_2 = 0$ we retrieve the state $\ket{\Psi}_{_{AB}}$ defined in Eq.~\eqref{Psi}. 

Since Alice's system is assumed to be described by  the  $d$-dimensional antisymmetric subspace of $d-1$  photons, the basis for Alice's system $\{\ket{A_i}\}$ can be written as:
\begin{align}
\ket{A_i}={1\over \sqrt{(d-1)!}}\det(\Lambda_{di})\ket{0},
\end{align}
where $\det(\Lambda_{di})$ are the minors of the matrix $\Lambda$ (cp. Eq.~\eqref{lambda}) with respect to the last row. Similarly we can define the basis $\{\ket{B_i}\} \equiv \{(-1)^ia_{i}^\dagger\ket{0}\}$ for Charlie's system. The index $i$ represents the different OAM modes of  the single photon. In this notation the states $|\Psi\rangle_{m_1m_2}$ read:
\begin{align}
|\Psi\rangle_{m_1m_2} &= \sum_{j} \exp\left(i\frac{2\pi}{d}jm_1\right) C_o(\Lambda)_{[dj]}a^{\dagger}_{d((j+ m_2)\hspace{-0.2cm}\mod d)}\ket{0}.
\end{align}
Here  $C_o(\Lambda)_{[di]} = (-1)^i\det(\Lambda_{di})$ is the $[di]^{th}$ cofactor of the matrix $\Lambda$.

The action of a $d$-input-$d$-output beam splitter setup, such as the Bell filter (cp. Fig.~\eqref{multi-bs}), can be represented by a unitary matrix $U$ that transforms the matrix $\Lambda$ as $U: \Lambda \to \tilde{\Lambda} = U\Lambda$. After passing through the beam splitter setup, the state $\ket{\Psi}_{m_1m_2}$ transforms to $\ket{\tilde\Psi}_{m_1m_2}$:
\begin{align}
|\tilde{\Psi}\rangle_{m_1m_2} &= \sum_{j\,k} \exp\left(i\frac{2\pi}{d}jm_1\right)U_{dk} a^{\dagger}_{k(j+ m_2\hspace{-0.2cm} \mod d)} C_o(\tilde{\Lambda})_{[dj]} \ket{0}. \label{eqn-23}
\end{align}
Since $\tilde{\Lambda} = U\Lambda$, we can use the following identity:
\begin{align}
C_o(U\Lambda)_{[ij]} &= \sum_{l}C_o(U)_{[il]}C_o(\Lambda)_{[lj]}
\end{align}
and rewrite Eq.~\eqref{eqn-23} as:
\begin{align}
|\tilde{\Psi}\rangle_{m_1m_2} &= \sum_{j\,k} \exp\left(i\frac{2\pi}{d}jm_1\right)U_{dk}a^{\dagger}_{k((j+ m_2) \hspace{-0.2cm}\mod d)} \sum_lC_o(U)_{[dl]}C_o(\Lambda)_{[lj]}\ket{0}.
\end{align}
We are interested only in those terms in the state $\ket{\tilde\Psi}_{m_1m_2}$ that have exactly one photon in each of the output ports and thus may lead to a coincidence count. Note that among the two indices of the creation operator $a^\dagger_{kj}$ the first index $k$ represents the spatial mode and the second index $j$ stands for the OAM mode. Furthermore, $C_o(\Lambda)_{[lj]}$ does not contain any creation operator in the spatial mode $l$. Therefore, the terms in the state $\ket{\tilde\Psi}_{m_1m_2}$ where $k=l$ contain exactly one photon in each spatial mode and hence may result in a coincidence count. After projecting the state $\ket{\tilde\Psi}_{m_1m_2}$ on the subspace of states that correspond to coincidence counts we get an unnormalized state $\ket{\Phi}_{m_1m_2}$ given by:
\begin{align}
\ket{\Phi}_{m_1m_2} &= \sum_{j\,k} \exp\left(i\frac{2\pi}{d}jm_1\right)U_{dk}a^{\dagger}_{k((j+m_2) \hspace{-0.2cm} \mod d)} C_o(U)_{[dk]}C_o(\Lambda)_{[kj]}\ket{0}.
\end{align}
In the above expression, if we choose 
\begin{align}
\eta_{dk} \equiv U_{dk} C_o(U)_{[dk]},\label{sufficient-cond}
\end{align}
i.e. $\eta_{dk}\equiv\eta$ to be $k$ independent then the expression reads:
\begin{align}
\ket{\Phi}_{m_1m_2} &= \eta\sum_{j\,k} \exp\left(i\frac{2\pi}{d}jm_1\right)a^{\dagger}_{k((j+ m_2) \hspace{-0.2cm} \mod d)}C_o(\Lambda)_{[kj]} \ket{0}
\end{align}
which is zero unless $m_1 = m_2 = 0$. Therefore, the requirement that $\eta$ is $k$ independent, is a sufficient condition for a beam splitter setup to project $\ket{\Phi}_{m_1m_2}$ to non-coincidence count states. 

\begin{figure}
\includegraphics[width=8.5cm]{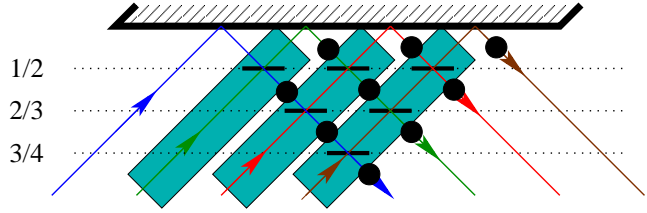}
\caption{A general scheme to project into and thus prepare  a $d$-photon totally antisymmetric state. In this scheme an array of beam-splitters (with indicated reflectivity) is used in a pyramidal structure. Each arm (turquoise box) consists of a Bell-filter as depicted in Fig.~3  of the main text and is followed by an array of single-photon heralding devices in each path.  A $d$-fold coincidence count in the output modes will ensure a projection on the antisymmetric state.}\label{general}
\end{figure}

Now we can prove that the unitary matrix $U_d$ corresponding to the Bell-Filter in Fig.\, \eqref{multi-bs} satisfies Eq.\,\eqref{sufficient-cond}. We can write the $U_d$ corresponding to $d-1$ beam-splitters as the product of $S_d$, the matrix for the $1/d:(d-1)/d$ beam-splitter  and $\tilde{U}_d$, which corresponds to the rest of the setup.
Therefore,
\begin{align}
S_d &= \left(\begin{array}{cccc}
\sqrt{\frac{d-1}{d}} & 0 & -\sqrt{\frac{1}{d}}\\
0 & \mathbb{I} & 0\\
\sqrt{\frac{1}{d}} & 0 & \sqrt{\frac{d-1}{d}}
\end{array}\right),\label{sd}\\
\tilde U_d &= \left(\begin{array}{cc}
1 &0\\
0 &U_{d-1}
\end{array}\right).\label{ud}
\end{align}
The sufficiency condition \eqref{sufficient-cond} then reads:
\begin{align}
\eta_{di} = (U_d)_{di}C_o(U_d)_{[di]} &= \sum_k
(S_d)_{dk}(\tilde{U}_{d})_{ki}\times \sum_l
C_o(S_d)_{[dl]}C_o(\tilde{U}_{d})_{[li]}.\label{eta}
\end{align}
From Eq.\,\eqref{sd} we have:
\begin{align}
(S_d)_{dk} &= -\sqrt{\frac{1}{d}}\delta_{k1} +
\sqrt{\frac{d-1}{d}}\delta_{kd},\quad C_o(S_d)_{[dl]} = -\sqrt{\frac{1}{d}}\delta_{l1} + \sqrt{\frac{d-1}{d}}\delta_{ld}.\label{eqn20}
\end{align}
Substituting \eqref{eqn20} in \eqref{eta}, we get:
\begin{align}
\eta_{di} =& \left(-\sqrt{\frac{1}{d}}(\tilde{U}_{d})_{1i} +
  \sqrt{\frac{d-1}{d}}(\tilde{U}_{d})_{di}\right)\times\left(-\sqrt{\frac{1}{d}}C_o(\tilde{U}_{d})_{[1i]}  + \sqrt{\frac{d-1}{d}}C_o(\tilde{U}_{d})_{[di]}\right)\label{eqn21}
\end{align}
It is easy to see that 
\begin{align}
(\tilde{U}_{d})_{1i} &= \delta_{i1},\quad
(\tilde{U}_{d})_{di} = (1-\delta_{i1})(U_{d-1})_{d-1\, i-1},\\
C_o(\tilde{U}_{d})_{[1i]} &= \delta_{i1},\quad
C_o(\tilde{U}_{d})_{[di]} = (1-\delta_{i1})C_o(U_{d-1})_{[d-1\, i-1]}.\label{eqn23}
\end{align}
Therefore, when $i = 1$ then only $(\tilde{U}_{d})_{11}$ and $C_o(\tilde{U}_{d})_{[11]}$ survives resulting in $\eta_{di} = 1/d$. When $i \ne 1$, then $(\tilde{U}_{d})_{11}$ and $C_o(\tilde{U}_{d})_{[11]}$ are zero. Thus, the only term in Eq.\,\eqref{eqn21} which survives is  $(d-1)(\tilde{U}_{d})_{di}C_o(\tilde{U}_{d})_{[di]} /d = (d-1)(U_{d-1})_{d-1\, i-1}C_o(U_{d-1})_{[d-1\, i-1]}/d = 1/d$. Here we used the relation  $(U_{d-1})_{d-1\, i-1}C_o(U_{d-1})_{[d-1\, i-1]} = 1/(d-1)$ which is assumed to hold for Bell-filter with $d-2$ beam-splitters. \\
\\
\noindent {\bf State preparation}\\
\noindent A general scheme to prepare (project into) an antisymmetric state  of $d$ photons can be realized  by an array of beam splitters together with non-linear optical devices \cite{Konrad2006, Invernizzi2005} that suppress the two- and higher-photon component in a path (cp.\ Fig.~\ref{general}). The scheme makes  repeated use of the Bell-filter shown in Fig.~3 in the main text. %\ref{multi-bs}.  
The latter can map the $(n-1)$-photon antisymmetric state to the $n$-photon antisymmetric state with the help of an additional photon in the case of $n$-fold coincidence counts in its output ports. Hence, if the output of the Bell-filter can be monitored to ensure that only one photon comes out of each of the output port, then this output can be fed to another such setup to generate an $(n+1)$-photon antisymmetric state. Therefore,  one can generate a $d$-photon totally-antisymmetric state, recursively, from a product state of $d$ photons. 

To prepare the totally antisymmetric state of three photonic qutrits (three-level systems)  a single Bell-filter (cp.\  Fig.~3) %\ref{multi-bs}
that consists of two beam splitters suffices. This requires as input a pair of photons in an antisymmetric state $\ket{\psi}=\ket{12}-\ket{21}$ and a single photon in state $\ket{3}$.  In addition, one needs to confirm that one photon leaves each output port. This can be achieved without single-photon heralding by sending two photons from different output ports to Bob who confirms this by a  state tomography and one photon  to Alice together with a photon pair from Charlie. A subsequent three-fold coincidence count after Alice's Bell filter (a copy of the one used for preparation) then announces successful state preparation as well as teleportation  of Charlie's qutrit which is encoded in the antisymmetric space of two photons:   
\begin{align*}
\ket{\psi}_A = \alpha\,(\ket{12}-\ket{21}) + \beta\,(\ket{13}-\ket{31}) + \gamma\, (\ket{23}-\ket{32})\,.
\end{align*}

%-------------------------------------------------------------------
% Bibliography
%-------------------------------------------------------------------
%\bibliographystyle{/home/goyal/Dropbox/nature}
%\bibliography{/home/goyal/Dropbox/ref_02}

\begin{thebibliography}{10}

\bibitem{Wootters1982}
Wootters, W.~K. and Zurek, W.~H.
\newblock A single quantum cannot be cloned.
\newblock {\em Nature}{ \bf 299}, 802--803 (1982).

\bibitem{Werner1998}
Werner, R.~F.
\newblock Optimal cloning of pure states.
\newblock {\em Phys. Rev. A}{ \bf 58}, 1827--1832 (1998).

\bibitem{Bennett1993}
Bennett, C.~H. {\em et al.}
\newblock Teleporting an unknown quantum state via dual classical and
  einstein-podolsky-rosen channels.
\newblock {\em Phys. Rev. Lett.}{ \bf 70}, 1895--1899  (1993).

\bibitem{Bell1964}
Bell, J.~S.
\newblock On the Einstein Podolsky Rosen paradox.
\newblock {\em Physics}{ \bf 1}, 195--200 (1964).

\bibitem{Brassard1998}
Brassard, G., Braunstein, S.~L., and Cleve, R.
\newblock Teleportation as a quantum computation.
\newblock {\em Physica D: Nonlinear Phenomena}{ \bf 120}, 43--47 (1998).

\bibitem{Gottesman1999}
Gottesman, D. and Chuang, I.~L.
\newblock Demonstrating the viability of universal quantum computation using
  teleportation and single-qubit operations.
\newblock {\em Nature}{ \bf 402}, 390--393 (1999).

\bibitem{Knill2001}
Knill, E., Laflamme, R., and Milbum, G.~J.
\newblock A scheme for efficient quantum computation with linear optics.
\newblock {\em Nature}{ \bf 409}, 46--52 (2001).

\bibitem{Duan2001}
Duan, L.-M., Lukin, M.~D., Cirac, J.~I., and Zoller, P.
\newblock Long-distance quantum communication with atomic ensembles and linear
  optics.
\newblock {\em Nature}{ \bf 414}, 413--418 (2001).

\bibitem{Briegel1998}
Briegel, H.-J., D\"ur, W., Cirac, J.~I., and Zoller, P.
\newblock Quantum repeaters: The role of imperfect local operations in quantum
  communication.
\newblock {\em Phys. Rev. Lett.}{ \bf 81}, 5932--5935  (1998).

\bibitem{Lvovsky2009}
A.~I.~Lvovsky, B. C.~Sanders, W.~T.
\newblock Optical quantum memory.
\newblock {\em Nature Photonics}{ \bf 3}, 706--714 (2009).

\bibitem{Bouwmeester1997}
Bouwmeester, D. {\em et al.}
\newblock Experimental quantum teleportation.
\newblock {\em Nature}{ \bf 390}, 575--579 (1997).

\bibitem{Boschi1998}
Boschi, D., Branca, S., De~Martini, F., Hardy, L., and Popescu, S.
\newblock Experimental realization of teleporting an unknown pure quantum state
  via dual classical and einstein-podolsky-rosen channels.
\newblock {\em Phys. Rev. Lett.}{ \bf 80}, 1121--1125   (1998).

\bibitem{Furusawa1998}
Furusawa, A. {\em et al.}
\newblock Unconditional quantum teleportation.
\newblock {\em Science}{ \bf 282}, 706--709 (1998).

\bibitem{Nielsen1998}
Nielsen, M.~A., Knill, E., and Laflamme, R.
\newblock Complete quantum teleportation using nuclear magnetic resonance.
\newblock {\em Nature}{ \bf 396}, 52--55 (1998).

\bibitem{Pan2001}
Pan, J.-W., Daniell, M., Gasparoni, S., Weihs, G., and Zeilinger, A.
\newblock Experimental demonstration of four-photon entanglement and
  high-fidelity teleportation.
\newblock {\em Phys. Rev. Lett.}{ \bf 86}, 4435--4438  (2001).

\bibitem{Jennewein2001}
Jennewein, T., Weihs, G., Pan, J.-W., and Zeilinger, A.
\newblock Experimental nonlocality proof of quantum teleportation and
  entanglement swapping.
\newblock {\em Phys. Rev. Lett.}{ \bf 88}, 017903  (2001).

\bibitem{Kim2001}
Kim, Y., Kulik, S., and Shih, Y.
\newblock Quantum teleportation of a polarization state with a complete bell
  state measurement.
\newblock {\em Phys. Rev. Lett.}{ \bf 86}, 1370--1373 (2001).

\bibitem{Lombardi2002}
Lombardi, E., Sciarrino, F., Popescu, S., and De~Martini, F.
\newblock Teleportation of a vacuum--one-photon qubit.
\newblock {\em Phys. Rev. lett.}{ \bf 88}, 70402 (2002).

\bibitem{Marcikic2003}
Marcikic, I., de~Riedmatten, H., Tittel, W., Zbinden, H., and Gisin, N.
\newblock Long-distance teleportation of qubits at telecommunication
  wavelengths.
\newblock {\em Nature}{ \bf 421}, 509--513 (2003).

\bibitem{Bowen2003}
Bowen, W.~P. {\em et al.}
\newblock Experimental investigation of continuous-variable quantum
  teleportation.
\newblock {\em Phys. Rev. A}{ \bf 67}, 032302  (2003).

\bibitem{Babichev2003}
Babichev, S.~A., Ries, J., and Lvovsky, A.~I.
\newblock Quantum scissors: Teleportation of single-mode optical states by
  means of a nonlocal single photon.
\newblock {\em Europhys. Lett:}{ \bf 64}, 1--7 (2003).

\bibitem{Ursin2004}
Ursin, R. {\em et al.}
\newblock Quantum teleportation across the danube.
\newblock {\em Nature}{ \bf 430}, 849 (2004).

\bibitem{Fattal2004}
Fattal, D., Diamanti, E., Inoue, K., and Yamamoto, Y.
\newblock Quantum teleportation with a quantum dot single photon source.
\newblock {\em Phys. Rev. Lett.}{ \bf 92}, 037904  (2004).

\bibitem{Riebe2004}
Riebe, M. {\em et al.}
\newblock Deterministic quantum teleportation with atoms.
\newblock {\em Nature}{ \bf 429}, 734--737 (2004).

\bibitem{Barrett2004}
Barrett, M.~D. {\em et al.}
\newblock Deterministic quantum teleportation of atomic qubits.
\newblock {\em Nature}{ \bf 429}, 737--739 (2004).

\bibitem{Olmschenk2009}
Olmschenk, S. {\em et al.}
\newblock Quantum teleportation between distant matter qubits.
\newblock {\em Science}{ \bf 323}, 486--489 (2009).

\bibitem{Amri2010}
Al-Amri, M., Evers, J., and Zubairy, M.~S.
\newblock Quantum teleportation of four-dimensional qudits.
\newblock {\em Phys. Rev. A}{ \bf 82}, 022329 (2010).

\bibitem{Ritter2012}
Ritter, S. {\em et al.}
\newblock An elementary quantum network of single atoms in optical cavities.
\newblock {\em Nature}{ \bf 484}, 195--200 (2012).


\bibitem{Hamadou.et.al13}
Hamadou Ibrahim, A., Roux, F.~S., McLaren, M., Konrad, T., and Forbes, A.
\newblock Orbital angular momentum entanglement in turbulence.
\newblock {\em Phys. Rev. A}{ \bf 88}, 012312 (2013).
 
\bibitem{Mafu.et.al13}
Mafu, M.{\em et al.} 
\newblock Higher-dimensional orbital-angular-momentum-based quantum key distribution
with mutually unbiased bases.
\newblock {\em Phys. Rev. A}{ \bf 88}, 032305 (2013).

\bibitem{Werner01}
Werner, R.~F.
\newblock All teleportation and dense coding schemes.
\newblock {\em J. Phys. A: Math and General}{ \bf 34}, 7081 (2001).

%%%%%%%%%%%%%%%%%%%%%%%%%%%%%%%
\bibitem{Mor1999}
Mor, T., and Horodecki, P.
\newblock Teleportation via generalized measurements, and conclusive teleportation.
\newblock arXiv:quant-ph/9906039, (1999).

\bibitem{Son2001}
Son, W., Lee, J., Kim, M.~S., and Park, Y.-J. 
\newblock Conclusive teleportation of a $d$ dimensional unknown state.
\newblock {\em Phys. Rev. A}{ \bf 64}, 064304 (2001).

\bibitem{Gu2002}
Gu, Y.-J., Zheng, Y.-Z., and  Guo, G.-C.
\newblock Conclusive teleportation and entanglement concentration.
\newblock {\em Phys. Lett. A}{ \bf 296}, 157--160 (2002).

\bibitem{Kim2004}
Kim, H.,  Cheong, Y.~W., and  Lee, H.-W.
\newblock Generalized measurement and conclusive teleportation with nonmaximal entanglement.
\newblock {\em Phys. Rev. A}{ \bf 70}, 012309 (2004).
%%%%%%%%%%%%%%%%%%%%%%%%%%%%%%%%%%%

\bibitem{Zeilinger07}
Ursin, R.
\newblock Entanglement-based quantum communication over 144 km
\newblock {\em Nature Physics}{ \bf 3}, 481--486 (2007).

\bibitem{Boyd}
Boyd, R.
\newblock {\em Nonlinear optics}
\newblock (Electronics \& Electrical. Academic Press, 2003).
 

\bibitem{Halevy2011}
Halevy, A., Megidish, E., Shacham, T., Dovrat, L., and Eisenberg, H.~S.
\newblock Projection of two biphoton qutrits onto a maximally entangled state.
\newblock {\em Phys. Rev. Lett.}{ \bf 106}, 130502 (2011).

\bibitem{Goyal2013}
Goyal, S.~K. and Konrad, T.
\newblock Teleporting photonic qudits using multimode quantum scissors.
\newblock {\em Scientific Reports }{ \bf 3}, 3548 (2013). 
 

\bibitem{Lutkenhaus1999}
L\"utkenhaus, N., Calsamiglia, J., and Suominen, K.-A.
\newblock Bell measurements for teleportation.
\newblock {\em Phys. Rev. A}{ \bf 59}, 3295--3300 (1999).

\bibitem{Hong1987}
Hong, C.~K., Ou, Z.~Y., and Mandel, L.
\newblock Measurement of subpicosecond time intervals between two photons by
  interference.
\newblock {\em Phys. Rev. Lett.}{ \bf 59}, 2044--2046 (1987).

\bibitem{Bose2002}
Bose, S. and Home, D.
\newblock Generic entanglement generation, quantum statistics, and
  complementarity.
\newblock {\em Phys. Rev. Lett.}{ \bf 88}, 050401 (2002).

\bibitem{Allen1992}
Allen, L., Beijersbergen, M.~W., Spreeuw, R. J.~C., and Woerdman, J.~P.
\newblock Orbital angular momentum of light and the transformation of
  laguerre-gaussian laser modes.
\newblock {\em Phys. Rev. A}{ \bf 45}, 8185--8189  (1992).

\bibitem{Konrad2006}
Konrad, T., Nock, M., Scherer, A., and Audretsch, J.
\newblock Production of heralded pure single photons from imperfect sources
  using cross-phase-modulation.
\newblock {\em Phys. Rev. A}{ \bf 74}, 032331 (2006).

\bibitem{Lancaster}
Lamcaster, P. and Tismenetsky, M.
\newblock {\em The theory of matrices}
\newblock (Harcourt Brace Jovanovich Publishers,  1985).

\bibitem{Invernizzi2005}
Invernizzi, C., Olivares, S., Paris, M. G.~A., and Banaszek, K.
\newblock Effect of noise and enhancement of nonlocality in on/off
  photodetection.
\newblock {\em Phys. Rev. A}{ \bf 72}, 042105  (2005).

\bibitem{Pegg1998}
Pegg, D.~T., Phillips, L.~S., and Barnett, S.~M.
\newblock Optical state truncation by projection synthesis.
\newblock {\em Phys. Rev. Lett.}{ \bf 81}, 1604--1606  (1998).


\end{thebibliography}

\makeatletter
\renewcommand\@biblabel[1]{#1.}
\makeatother
%\bibliographystyle{/home/goyal/Dropbox/Science}

%==========================================================

\section*{Acknowledgements} We thank A.\ Forbes, P.\ Krumm, K.\ Garapo and J.\ Leach for useful discussions.  T.K. acknowledges  partial support from the National Research Foundation of South Africa (Grant No. 86325 (UID)).

\section*{Author contributions}
P.E.B.-D. carried out preliminary investigations. S.K.G. and T.K. solved the problem. F.S.R. and S.G. provided mathematical insights. S.K.G., S.G., F.S.R. and T.K. wrote the manuscript. All authors reviewed the manuscript.

\end{document}